%% file: main.tex
\documentclass[10pt, conference,compsocconf]{setup/IEEEtran}

\input{setup/packages}

\input{setup/macros}

\input{setup/table}

\begin{document}

\title{Buy your Coffee with Bitcoin: Real-World Deployment of a Bitcoin Point of Sale Terminal} 

\author{
\IEEEauthorblockN{Shayan Eskandari} 
\IEEEauthorblockA{BitAccess}\and
\IEEEauthorblockN{Jeremy Clark} 
\IEEEauthorblockA{Concordia University}\and
\IEEEauthorblockN{Abdelwahab Hamou-Lhadj} 
\IEEEauthorblockA{Concordia University}
}

\maketitle


\input{sections/00-abstract}


\input{sections/11-body}


\bibliographystyle{IEEEtranS}
\footnotesize
\bibliography{bib/bib,bib/bitcoin}
\normalsize


\end{document}

%% file: setup/packages.tex
\usepackage{graphicx}
\graphicspath{{figures/}}
\DeclareGraphicsExtensions{.jpg,.png}

\usepackage{bibspacing}
\usepackage[caption=false,font=footnotesize]{subfig}

\usepackage{amsmath}
\usepackage{stmaryrd}

\usepackage{array}
\usepackage{color}
\usepackage[hyphens]{url}
\usepackage[pdftitle=A\ First\ Look\ at\ the\ Usability\ of\ Bitcoin\ Key\ Management,pdfauthor=Shayan\ Eskandari\ et\ al.]{hyperref}
\usepackage{comment}
\usepackage{cite}


%% file: setup/macros.tex
\usepackage{xspace}

\newcommand{\eg}{\textit{e.g.,}\xspace}

\renewcommand\paragraph[1]{\smallskip\noindent\textbf{#1.}}

\newcommand{\Aunja}{\textsf{Aunja PoS}\xspace}


%% file: setup/table.tex
\usepackage{adjustbox}
\newcommand{\headrow}[1]{\multicolumn{1}{c}{\adjustbox{angle=45,lap=\width-0.5em}{#1}}}

\newcommand{\full}{$\bullet$}
\newcommand{\prt}{$\circ$}

%% file: sections/00-abstract.tex

\begin{abstract}
In this paper we discuss existing approaches for Bitcoin payments, as suitable for a small business for small-value transactions. We develop an evaluation framework utilizing security, usability and deployability criteria, and examine several existing systems and tools. Following a requirements engineering approach, we designed and implemented a new Point of Sale (PoS) system that satisfies an optimal set of criteria within our evaluation framework. Our open source system, Aunja PoS, has been deployed in a real world caf\'{e} since October 2014.
\end{abstract}


%% file: sections/11-body.tex



\section{Introductory Remarks}

Bitcoin is a cryptographic currency publicly proposed in 2008~\cite{Nak08}. It has reached a level of adoption unrealized by decades of previously proposed digital currencies (from 1982~\cite{Cha82} onward). Unlike most previous proposals, Bitcoin does not distribute digital monetary units to users. Instead, a public ledger (called the blockchain) maintains a list of every transaction\footnote{Technically, a transaction specifies a short script that encodes how the balance can be claimed as the input to some future transaction.} made by all Bitcoin users since the deployment of the currency in January 2009. 

While Bitcoin was originally envisioned for online currencies, a number of businesses have begun to accept Bitcoin in person. To our knowledge, the academic community has not given any attention to Bitcoin point-of-sale (PoS) terminals and their unique requirements in terms of security, usability, and deployability. In this paper, we develop a framework for evaluating competing approaches. We then provide a case study detailing the design and implementation of our open source PoS terminal, \Aunja, which was made for a caf\'{e} in Montreal\footnote{Cafe Aunja \url{http://aunja.com}} following the SCRAM requirements engineering approach\footnote{Scenario-based Requirements Analysis Method} (see Section~\ref{SCRAM}) and has been in operation since 2014. 
  
\section{Evaluation Framework}

We propose a framework for comparing Bitcoin PoS solutions, scoring the competing systems on usability, deployability, and security (following~\cite{BHOS12}). These are not a full set of requirements for a general purpose PoS system but are tailored to in-person low volume transactions that you might find in a small business. The requirements are adapted from our previous framework for Bitcoin wallets~\cite{eskandari2015first} and originated with new requirements based on our expertise. The requirements are used to score each system in Table ~\ref{tab:method-comp}. For simplifying the figure, we use three score indicators. (\full) for a complete score on the requirement, (\prt) if the requirements has not met completely and empty space if it is not satisfying the need. For some of the requirements the scoring system might not be intuitive (\eg low cost to run) however we justify each score later in the paper.

\subsection{Usability} We consider the following aspects of usability.

\begin{itemize}

\item \textbf{User-Friendly:} This is a general category to note any usability violations that would result in a payment process being too technical or complex for the employee or customer. A single training session for the employee should suffice and the system should be intuitive to a Bitcoin user. There should be a clear and mutual understanding when the payment is finalized. A PoS that has all of these features would score (\full), having some would result in (\prt).

\item \textbf{Time-Efficient: }Processing payments should not take significantly more time than common payment systems such as credit card payments. If the process takes the same time as credit card payments it would score (\full), anything more than that would be (\prt) or none.

\item \textbf{Fair Exchange Rate: } There should be an easy and verifiable approach for the payer and payee to come to a consensus on fiat currency to Bitcoin exchange rate. If the price is retrieved from commonly accepted sources it would score (\full).

\item \textbf{Availability: }All employees should be able to do the Bitcoin payment process without the need to know any credentials. If it is on a public domain for anyone to access it will score (\full), if it needs some private information it will score (\prt) and if it needs credentials it will score none.

\end{itemize}
\subsection{Deployability} We use this category to state the requirements regarding implementation.

\begin{itemize}

\item \textbf{Low Cost to Run: }PoS should be accessible with one of the currently owned devices of the caf\'{e} such as the cashier computer, the PoS terminal\footnote{The common PoS that accepts Visa/Debit Cards} or mobile devices. There should not be a need for buying new hardware or expensive software. For this requirement, we would score a (\full) to a free of monetary cost system, and a (\prt) score to a moderate amount of spendings.

\item \textbf{Enables Branching: }The ability to install the point of sale on multiple branches of the business. Configuration might be needed to differentiate two branches in the system. If the PoS is packaged and easy to install on the second branch of the business it will score (\full), if it needs some modification (\prt) and if it is the same procedure to install it as the first one it will score none.

\end{itemize}
 
\subsection{Privacy} As transactions in Bitcoin are published to the blockchain, it is important to consider both payer or payee privacy.
\begin{itemize}

\item \textbf{No Information leakage: } There should no sensitive information available to the customer when she wants to pay with Bitcoin. This information might include the infrastructure of the business's network or a private domain used for accounting purposes. If it leaks any sensitive information it will score none and if it leaks some non-sensitive information it will score (\prt). 

\item \textbf{Maintains Payee's Privacy: }The payer should not be able to see how much the payee has received prior or after her payment but just her own amount of payment. If there is no link between the payments visible to the payer the PoS will score (\full).

\item \textbf{Maintains Payer's Privacy: }The payee should not be able to see how much the payer owns. Note that this challenge has not been fully solved (\textit{c.f.,} ~\cite{androulaki2013evaluating}). All the PoS in this evaluation scored (\prt), including our own, but we include this property to have a complete framework for the evaluation of future software that may utilize privacy-preserving add-ons~\cite{BNMC+14} or cryptocurrencies~\cite{MGGR13}.

\item \textbf{Confidential Payments List: }The ability to see the payments list, only available for the manager by an authentication method, such as a password-protected panel. If the PoS offers a report page for the manager it will score (\prt), if the report page could have hierarchal authentication for employees with limited access it will score (\full). 

\end{itemize}
\subsection{Security} Security  is one of the most important aspects in any financial payment system. Security of the system represents more than just the PoS code, it includes the environment which PoS is being used, the people using the software and the operating environment of the software.
\begin{itemize}

\item \textbf{No 3rd-Party Trust: }There should be as little third party trust as possible to accept and hold Bitcoin. Full trust to a third party will result in scoring none, some trust on the main functionality of the PoS result in (\prt) and no trust will result in (\full) score.

\item \textbf{Data Encryption: }In the case of any attacks on the service, there should be security measures that makes sure the attacker will not be able to have access to the private keys and transfer Bitcoin funds. Only if all the sensitive data is encrypted, the PoS will score (\full).

\item \textbf{No Software Dependency: }The system should use as little dependencies as possible to minimize the attack vector on the server. If the PoS needs complex set of software or hardware to work, it will score none, and if it could be executed in a browser\footnote{In order to use a software PoS a mobile device or a computer is needed and we assume a web browser is by default installed on these devces} without the need to run any other software it will score (\full).

\end{itemize}

\section{Evolution of PoS proposals}

\input{sections/newtable}

Most existing payment systems suit the online markets (\eg e-commerece) and not physical points of sale.\footnote{\url{https://en.Bitcoin.it/wiki/How_to_accept_Bitcoin,_for_small_businesses}} We list all the available approaches to accept Bitcoin payments that can at least be adapted for in-person transactions.

\subsection{Single Bitcoin address displayed} 
A simple way for small businesses to accept Bitcoin is to generate one Bitcoin address and display it, say as a QR code. Customers can scan the QR code and input the dollar value on their Bitcoin wallet and pay the business with the equivalent Bitcoins. \\

\textbf{Usability:} This approach puts the employee in a position to prepare, receive and check Bitcoin payments manually (User friendly: none). This makes the time spent on the payment longer than an integrated payment system (Time-efficient: none). Price conversion from the local currency to BTC would also be a manual lookup (Fair exchange rate: none). Technical training is required for each employee responsible for handling Bitcoin payments. As long as the QR-code print is visible to the payer, it is available to pay (availability: \full).

\textbf{Deployability:} The cost to implement this method is almost zero (Low cost to run: \full). In case there are multiple branches, more print outs suffice to have multiple point of sales (Enables branching: \prt).

\textbf{Privacy:} This method provides no privacy for the seller (Payee's privacy: none). As all the Bitcoin transactions are publicly available in the Blockchain, anyone with the knowledge of the Bitcoin address could see all the received payments, thus anyone could have access to the reporting page (Confidential Payments list: none).

\textbf{Security:} Other than the system holding the private key, security does not factor into this approach (No 3rd-party trust: \full). The private key should be kept in a secure place, preferably a cold storage unless the funds should be transferred to another address (\eg to exchange for cash). There are no software or data involved thus there is no software dependency (Data Encryption: none, No software dependency: \full).

\subsection{Hardware terminals}
There are multiple hardware terminals proposed for accepting Bitcoin\footnote{Bitstraat \url{bitstraat.nl}, Xbterminal \url{xbterminal.com} , Coinkite \url{coinkite.com}}, however due to the high cost to run (\eg Coinkite\footnote{\url{https://coinkite.com/store/products/all}} PoS are for sale at the starting price of 970USD), they have not been used in most of the small businesses and have not been reviewed before. At the time of writing all the proposed hardware terminals are unavailable to purchase and the future of Bitcoin hardware terminals is indeterminate.

\textbf{Usability:}
The interfaces of each of the provided terminals are different. The most popular ones mimic the look and feel of a normal point of sale terminal used by credit card companies. However adding a new device to the payment routine would make it less user friendly and arises the need for training the employees (User friendly: \prt). The time and availability of the payment through a hardware terminal should be the same as credit card payments if not lower (Time-efficient: \full) . The customer, nor the payee has any control over the exchange rate and it is provided by the PoS terminal operator (Fair exchange rate: \prt). The device is accessible to anyone who has access to the other payment terminals (Availability: \full).

\textbf{Deployability:}
Due to the high costs, they score low in our framework (Low cost to run: none). Also in case there are multiple branches of the business, there should be one devices bought for each branch this makes the costs even higher (Enables branching: none).

\textbf{Privacy:}
Accepting Bitcoin with a hardware terminal should persevere the privacy the same as the regular credit card terminals, however the payees privacy depends on the implementation of the Bitcoin payment system (Payee's privacy: \full). The terminal providers also offer similar interface to credit card terminals to list the payments (Confidential Payments list: \full).

\textbf{Security:}
The payee has no control over his private keys nor holds the funds (No 3rd-party trust: none), thus he needs to trust the third-party company that provided the terminals to keep the funds safe, and will receive the payments upon the agreed time frame with probably small transaction fees. As for other aspects of security, we assume the back-end implementation keeps the private keys encrypted and secure (Data encryption: \full). There are security risks involved in adding new hardware or software to the cashier's computer that will fall out of the scope of this paper (No software dependency: none).

\subsection{Online Merchant Services}
Most of these services do not have an explicit implementation for a physical payment system. 
Two popular ones, at the time of writing, are Bitpay\footnote{\url{https://bitpay.com}}  (0\% fees) and Coinbase\footnote{\url{http://coinbase.com}} (1\% on exchanging Bitcoins to fiat currency). 

 \textbf{Usability:}
Implementing a Bitpay payment is straightforward and easy to implement. There are not many jargon or technical options for the employee (User friendly: \full). They have their own exchange rate (Fair exchange rate: \prt) that the business owner could set to exchange to cash as soon as he receives payments, this will remove the possible effect that Bitcoin price volatility could have on the payments. It requires some credentials to access the PoS page (Availability: \prt).

 \textbf{Deployability:}
The only thing required by this approach is a smart phone or a small computer that users could interact with and browse to the Bitpay payment page, preferably with a touchscreen for easier price input and user interaction (Low cost to run: \full). It is easy to add more branches to the original account or even make a new account for the second branch (Enables branching: \full).

 \textbf{Privacy:}
Bitpay has another approach for preserving the privacy. As they generate a new address for each transaction, the payee's privacy is safe(Payee's privacy: \full). However there has been reports of account suspensions because the payments were coming from flagged Bitcoin addresses (\eg black markets\footnote{Darknet Blackmarkets\url{https://en.wikipedia.org/wiki/Darknet_market}} or LocalBitcoins \footnote{Peer to peer Bitcoin trading site \url{http://localBitcoins.com}}). In this case, the privacy, as the sense that we are evaluating, is being held but maybe not in all aspects needed in a payment system. In order to view the payments, business owner should log in his account and view the payments but other employees cannot see the list using any other accounts (Confidential Payments list: \prt)

 \textbf{Security:}
Every aspect of the payment system is implemented by Bitpay, they offer one of the most secure payment systems so far and there has been no big hacks reported (Data encryption: \prt) . However, user has no control over his private keys and all the keys are being stored on Bitpay servers (No 3rd-party trust: none) which means complete trust to a third party. As they are a web-based solution, a device with a browser is enough to use their PoS (No software dependency: \full)

\subsection{Mycelium Gear}
Mycelium Gear \footnote{\url{https://gear.mycelium.com/}} is a service offered by the Mycelium group that offers a widget as an interface to the user and a service that would use the BIP32\footnote{Hierarchical Deterministic Wallets}~\cite{bip32proposal}  public key provided on the Admin panel to generate new addresses securely. This means that they don't hold any private keys, but still uses the same set of paths for address generation as their Mycelium Mobile wallet uses.

\begin{figure}[htb!p]
\centering
\includegraphics[scale=0.4]{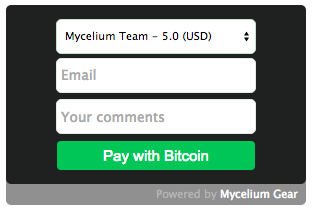}
  \caption{Mycelium Gear Widget}
\label{fig:mycelium-widget}
\end{figure}

\textbf{Usability:}
Mycelium Gear is designed for e-commerce business and should be customized to suit a physical business PoS (User friendly: \prt) . There are no fees related to using this service, and they offer fast verifications on 0-confirmation transactions (Time-efficient: \full) and it's possible to chose from a list of supported exchanges to retrieve the Bitcoin exchange rate from (Fair exchange rate: \prt). A unique URL is needed to access the payment page and the employees should be aware of this link (Availability: \prt).

\textbf{Deployability:}
This method would be simple to implement but somehow more complicated to customize as there's not that much access to the code to be able to customize for business needs. Although the cost-to-run depending on the implementation could be almost zero (Low cost to run: \prt). The only deployability downside is that the payee is forced to use Mycelium Mobile wallet to manage his payments, however doing so makes it easy to use the PoS in other branches and dedicate different accounts to each branch (Enables branching: \full).

 \textbf{Privacy:}
As Mycelium Gear uses BIP32 to generate a new address for each transaction request the payees privacy is held (Payee's privacy: \full). However, there is no user management for the report page, If the customer closes the successful payment page, the employee would not be able to check if the payment was received or not unless he has the administrator password to check the transaction list (Confidential Payments list: \prt).

 \textbf{Security:}
Nothing related to the PoS holds any private information or keys that might be in danger of exposure, however all other aspects of the system is running on their infrastructure (No 3rd-party trust: \prt) . Although all the private keys would be in the Mycelium mobile wallet (No software dependency: \prt) that is not prone to mobile malwares or hardware failure (Data Encryption: \prt).


\subsection{Discussion}
As seen in Table \ref{tab:method-comp}, there is no perfect solution out of the box for a small business to start accepting Bitcoin. After further discussions with the business owner, we decided to implement our own custom PoS using available open source software. This way it would be easy to incrementally change the PoS system with the customer and employees feedback to meet the needs of the business. In the following sections, we describe \Aunja.

\section {Requirements Engineering}
Requirement engineering is a subfield of software engineering devoted to the pre-implementation process of software design, focusing on eliciting requirements from stakeholders, negotiating a balanced approach, and producing a system specification. RE originated in 1979~\cite{alford1979software} and was popularized about a decade later~\cite{dorfman1990system}.

\subsection{SCRAM}
\label{SCRAM}
We adapted SCRAM (Scenario-based Requirements Analysis Method)~\cite{REScenario} as a framework to gather the requirements of this system, as there are finite scenarios for payments in a small business. Scenarios are examples of real world experiences that we  use to model what is required from the system. SCRAM defines the following four phases of requirement engineering:

\begin{itemize}

\item \textbf{Initial requirements capture and domain familiarisation: } This is done by interviewing and fact-finding to gain a full understanding of how the business works.

\item \textbf{Storyboarding and design visioning: } This is done by creating walkthroughs to show to the business and get feedback on feasibility.

\item \textbf{Requirement exploration: } This uses the early prototypes  and designs to get feedback from the business and validate the requirements.

\item \textbf{Prototyping and requirement validation: } This is done by developing fully functional prototypes and continues refining the requirements until the product is acceptable to the business.

\end{itemize}

\textit{Phase 1:} We asked the caf\'{e} owner, two employees and two customers for a scenario involving Bitcoin payment to create the common ``normal use case.'' Exceptions to the normal use case could be something like a power failure, however this would also fail for current methods such as credit cards. As the caf\'{e} already has other payment systems in place, there proved to be no need to go through the caf\'{e}'s business plan or any other specifications to check for conflicts. Our only change is to implement an additional payment system at the cashier's desk. However there are Bitcoin-specific requirements like realtime Bitcoin exchange rates and obvious alert of successful or failed payments.

%

    \textit{Phase 2:}
    Based on the information gathered from Phase 1 and further analysis, such as user survey on the design, a storyboard was developed. 
%

\textit{Phase 3:}
We developed a ``concept demonstrator''~\cite{REScenario} capable of doing a simple Bitcoin payment. The Bitcoin exchange rate and transaction amount was hard coded and the transaction would be executed manually. We asked the employees to run a mock purchase with the demonstrator to see how they would interact with the system. As Bitcoin concepts might be ambiguous for the new user, there should not be any interactions with Bitcoin concepts and terminology. After the transaction was done, the owner pointed out that there is a need for a central logging system that could be checked from time to time for accounting purposes.

\textit{Phase 4:}
We used the feedback gathered from phase 3 to make the first prototype. The prototype retrieved the Bitcoin exchange rate in realtime and the employee only had to input the dollar amount in \Aunja. This made it possible to keep the Bitcoin terminology out of the scope of the training for the employees. However on the first prototype, to show the successful payments, the system was showing the transaction on Blockchain Explorer\footnote{\url{http://blockchain.info}} using web-based APIs. This was not clear enough for a novice user to determine the state of the transaction. On the second round of prototyping, we designed an interface to show that the transaction has been broadcasted to the Bitcoin network (called a 0-conf transaction; security discussed below).  
\section{Design and Implementation}
\label{Design and Implementation}

Multiple approaches for implementing \Aunja were apparent. One of the lower cost methods would be to use a computer on the caf\'{e}'s network as the web server however maintenance and support could be a difficult task. The network might be overwhelmed by the high number of connected devices and might not function properly. Uptime is one of the most important aspects for a payment system. The next low cost solution is to use shared hosting to host the wallet server and design a web based payment interface for the employees, including a secure reporting page. We opted for this approach and naturally chose to implement \Aunja in PHP, a popular language for shared hosting. 

\subsection{Implementation measurements}
\label{Implementation measurements}
After multiple rounds of surveying employees and customers to understand their needs and also researching the subject, here is the break down of the results:
\subsubsection{Usability} 
\begin{itemize}

\item \textbf{User Friendly (\full): } The interface should be minimal and simple, with the ability to show the exchange rate of Bitcoin to fiat, input box for the price in dollars, estimation of Bitcoin amount equivalent to the price and a note section.
As for the user facing interface, it should be simple, showing all the required information such as Bitcoin amount, the exchange rate and the QRCode for the deposit Bitcoin address. Both interfaces should indicate when the transaction is complete.

\item \textbf{Time-Efficient (\full): } It should not take more than normal payment system to initiate the payment. A web based interface would have the advantage that it can be loaded from any device with Internet access. Also to verify the payment it should not take longer than needed. It also needs to use fast verification methods to indicate that the payment is propagated to Bitcoin network. Knowing that a propagated transaction is not same as confirmed transaction but is an accepted risk for low volume transactions.

\item \textbf{Fair Exchange Rate (\full): } After some research, we opted for an HTTPS-enabled webservice called Bitcoinaverage\footnote{\url{https://bitcoinaverage.com}} which offers a transparent aggregation of various exchange rates to produce a fair spot price. 

\item \textbf{Availability (\full): } The payment interface should be open to public and should be loaded on any device.

\end{itemize}
\subsubsection{Deployability}
\begin{itemize}

\item \textbf{Low Cost to Run (\prt): } The only costs associated with this implementation would be the annual cost of the shared hosting that is less than \$100 for an unlimited web host. For the sake of this research, there would be no other implementation and development costs.

\item \textbf{Enables Branching (\prt): } For now there's no plan to have more branches for this business, but depending on the implementation, launching additional branches would only involve running additional instances of the application on the server.

\end{itemize}
 
\subsubsection{Privacy} 
\begin{itemize}

\item \textbf{No Information leakage: } The payment interface does not reveal any information about the backend nor the business' internal infrastructure.

\item \textbf{Maintains Payee's Privacy (\full): } There should be a new address generated for each transaction request so no one can see how much the business have received in Bitcoin prior or after each transaction.

\item \textbf{Maintains Payer's Privacy (\prt) : } This would be the payers Bitcoin wallet client responsibility and it would be out of the scope of this PoS system.

\item \textbf{Confidential Payments list (\full): } A reporting and administration interface is made accessible to the business owner or designated personals.

\end{itemize}
\subsubsection{Security} 
\begin{itemize}

\item \textbf{No 3rd-Party Trust (\full): } There should not be any sensitive usage of 3rd parties in the system, it should work as a stand alone system. Note that while a trusted party is referenced for the exchange rate, the received value is treated as an assertion to be verified.  

\item \textbf{Data Encryption (\full): } All the private keys should be encrypted and then stored on the server. 

\item \textbf{No Software dependency (\prt): } There should not be any software dependency on the payment page for the business. The software dependencies on the server side should all be included in the package as open source software.
\end{itemize}


\subsection{Open source libraries and software applications}
After the requirement engineering phase, we looked for PHP components to form the following base.

\begin{itemize}
\item \textbf{Bitcoin SCI: }Bitcoin Shopping Card Interface. 
\item \textbf{PHP Elliptic Curve library\footnote{\url{http://matejdanter.com}}: } Used as a dependency to Bitcoin SCI to generate Bitcoin addresses.
\item \textbf{Bitcoin-prices\footnote{\url{https://github.com/miohtama/Bitcoin-prices}}: } Display Bitcoin prices in human-friendly manner in fiat currency using Bitcoinaverage.com market data
\item \textbf{Bitcoin SCI } (Bitcoin Shopping Cart Interface \footnote{\url{http://bitfreak.info/?page=tools&t=bitsci}}): is a set of libraries and tools that enables the user to process Bitcoin transactions with only PHP. It is originally designed to be integrated in e-commerce websites but it could be easily modified to meet our needs.
\end{itemize}

\begin{figure}[htb!p]
\centering
\includegraphics[scale=0.5]{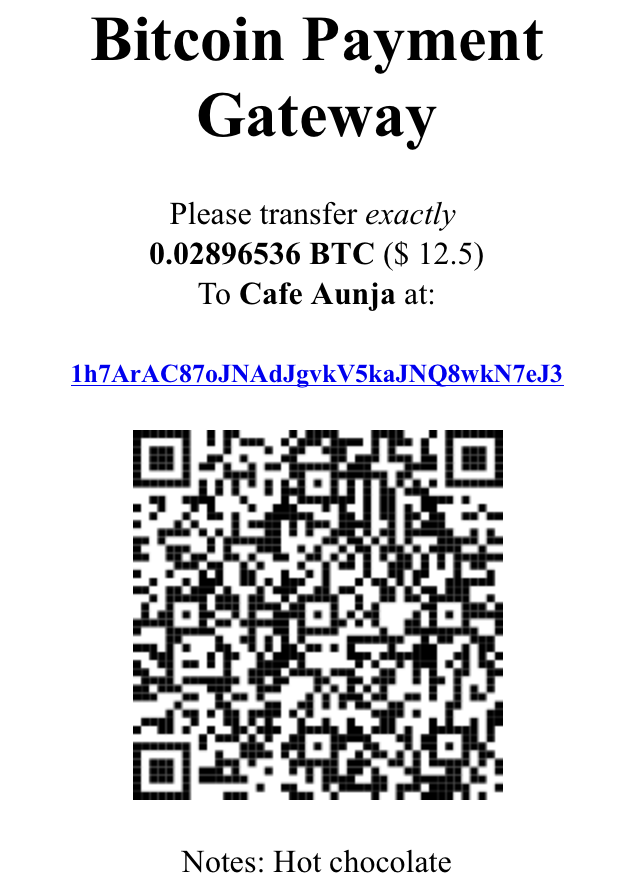}
  \caption{Bitcoin SCI (Bitcoin Shopping Cart Interface)}
\label{fig:Bitcoin-sci}
\end{figure}

The latter is not a complete project to process payments. The first decision was to use this package for building the prototype and then if we failed to modify the package to meet our needs, use another approach, however we did make it suit the needs and Bitcoin SCI was used in the end product. 

A break down of the tools Bitcoin SCI provides us are as follow:
\begin{itemize}
\item \textbf{Bitcoin Address generation: } Bitcoin SCI uses PHP Elliptic Curve library to generate new secure Bitcoin addresses (set of public and private keys)
\item \textbf{Private key encryption: } Using phpseclib\footnote{\url{http://phpseclib.sourceforge.net}}, all the private information (Bitcoin private keys, transaction details) are stored encrypted
\item \textbf{Payment Confirmation: } It uses APIs from a web tool\footnote{blockexplorer.com} to confirm receiving payments.
\item \textbf{Input Interface: } Even though this package was meant to be used as an e-commerce payment system, it has the basic tools and methods to build the price input page.
\end{itemize}

However it lacks some other features that should be added:

\begin{itemize}
\item \textbf{Database: }The management and report page requires saving the transaction details into a database.
\item \textbf{Fair Bitcoin Exchange rate: } It uses a predefined source to obtain the exchange rate of Bitcoin and it is not possible to set different Fiat currencies.
\item \textbf{User-Friendly interface: } All the interfaces are poorly designed and need to be modified to suit the PoS system.
\item \textbf {Report Page: } The report page requires authentication.
\item \textbf {Input Validation: } Other than security perspective of input validation, this is needed because of the way we want the PoS to work. It should alert the employee if she has done something wrong before going to the next page.
\item \textbf {Cash out option: } As all the private keys are stored encrypted in the server, we need a way to cash out the available Bitcoins and send them to another Bitcoin address. It is possible to retrieve the private keys of each Bitcoin address separately from the tool, but it's not scalable to multiple weekly transactions.
\end{itemize}

We use Bitcoin-prices to set Bitcoinaverage.com prices as our main source of price conversion, and it gives nice tools for interface design, such as the ability to switch between different currencies by just clicking on the price. This allows anyone deploying the system to reach a fair exchange rate in many different currencies.

We use Sweet Alert\footnote{\url{http://t4t5.github.io/sweetalert}} to facilitate user-friendly javascript alert messages. In the case of data validation, we needed a simple way to inform the employee that there is a mistake to be fixed. For this case, browser-based Javascript validation saves a roundtrip to the server.

\subsection{Prototyping}
With the full knowledge of the requirements and a few sketches of the interface, we started developing \Aunja. Although the first prototype was ready to launch within a week, we did 3 prototypes in the month after, as each had bugs fixed and features added as we surveyed and obtained feedback from the employees on each round of prototyping. 
\subsubsection{PoS main functionalities}
The PoS was hosted on a shared hosting service named Host Monster \footnote{\url{http://hostmonster.com}}. They offer low cost annual plans that offer PHP and MySQL which are the requirements that we needed.
Then we started working with Bitcoin SCI to add the database functionality and defined tables for transaction requests and payments on MySQL. Other tasks were involved in integrating the above mentioned open source projects into each other to have a complete solution package.

One of the features added on the second round of prototyping was the ability to show the Bitcoin price in USD other than the default CAD. This was added with the usage of the Bitcoin-prices library. It was possible to implement a drop down menu with all the caf\'{e}'s menu options to be added to the list but as we discussed this solution with the caf\'{e} owner, he mentioned that the items in the menu might not stay the same during the year and there might also be price changes, so that approach was not suitable for this business, although it might be a good option for an e-commerce site.

\subsubsection{Private reporting page}
One other aspect of the requirements was a reporting page. This was based on the feedbacks from the caf\'{e}'s owner and his preferences.

\begin{figure*}[ht] 
\centering
\includegraphics[width=\linewidth]{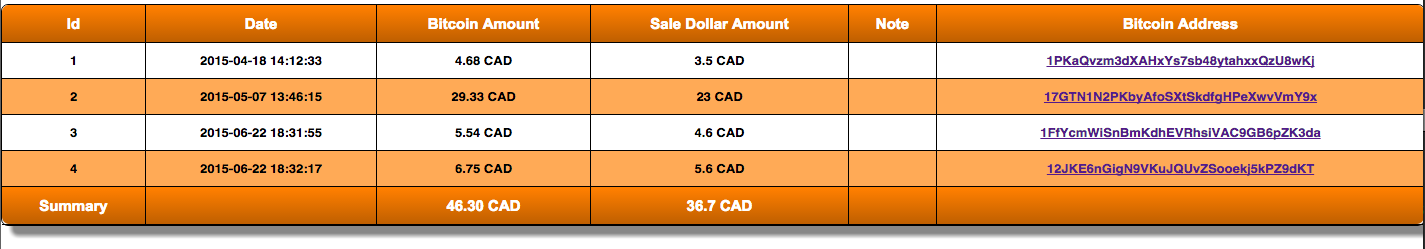}
  \caption{Report Page}
\label{fig:report_page}
\end{figure*}

One of the important fields added later to the report page was the ``Sale Dollar Amount.'' Bitcoin's price is volatile compared to other currencies and the caf\'{e} owner did not want to risk losing money by accepting Bitcoin. So as an agreement, we decided to lock the price of each sale on the sale time, as if he was selling his products with cash.\footnote{this method is actually one of the common methods recently used by Bitcoin payment processors.} Thus on the second prototype of the report page, this field was added for accounting purposes. Another feature request was the ability to check each transaction on a blockchain explorer and also decrypt and export the private keys of those addresses that has some balance. This has been done for the admin report page.

\Aunja has been made open source and available to public\footnote{\url{https://github.com/shayanb/Bitcoin-PoS-PHP}} under GNU General Public License v2 and has already been used in some other small businesses.

\subsection{Training}

\begin{figure}[htb!p]
\centering
\includegraphics[scale=0.1]{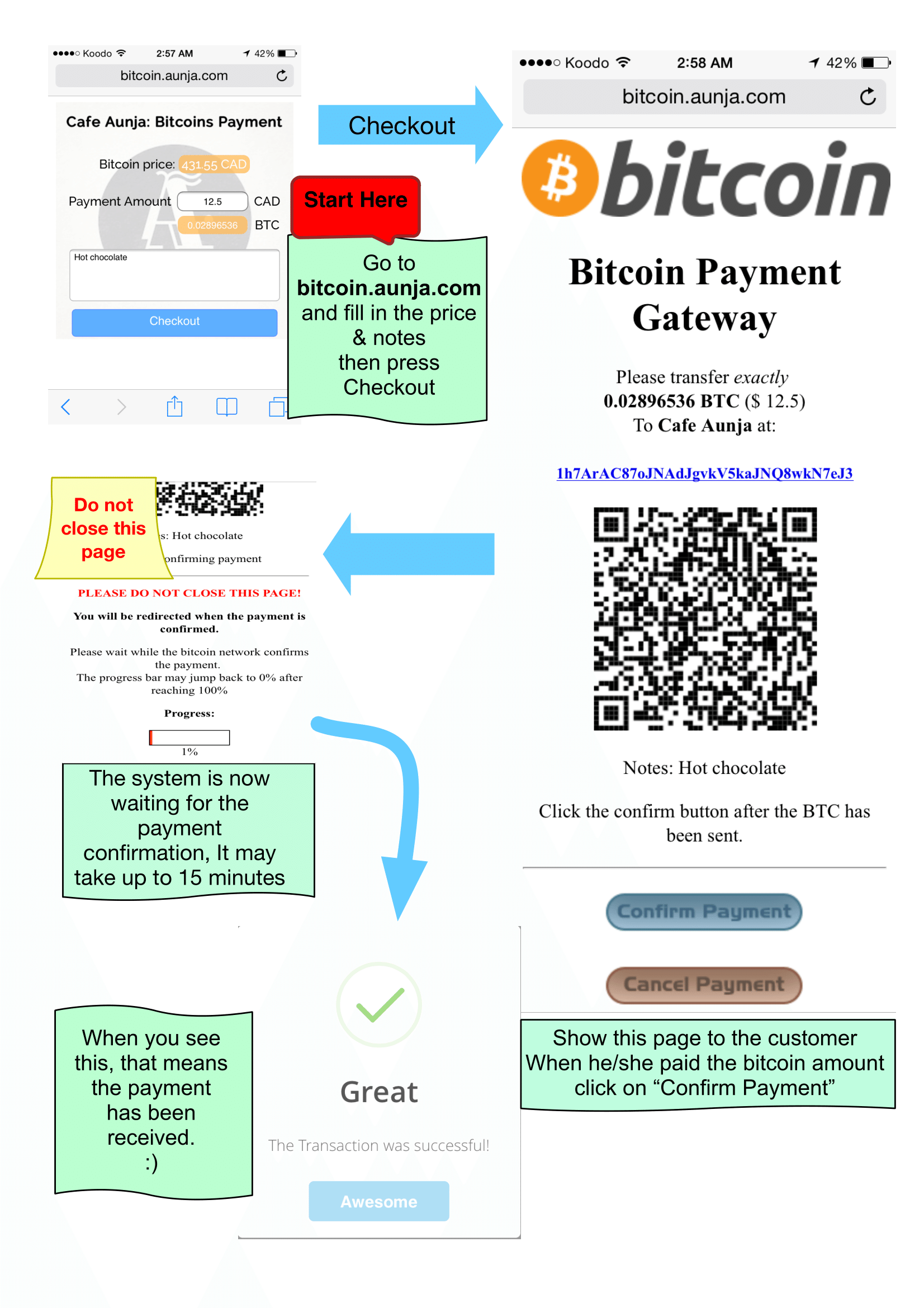}
  \caption{PoS - Step by step manual for Bitcoin payments}
\label{fig:payment_manual}
\end{figure}

There is no jargon or technical requirements to use \Aunja, but some details specific to Bitcoin transactions have to be taught to the employees to be able to recover from human errors while a transaction is being processed. Other than in-person training that was done with every employee, a manual was made (Figure~\ref{fig:payment_manual}) and was attached to the cashier's counter for future reference by all caf\'{e} employees.

\section{Real-world Deployment}
Caf\'{e} Aunja started accepting Bitcoin with \Aunja on Oct 23, 2014, and by our knowledge was the first caf\'{e} in eastern Canada that accepts Bitcoin.

\subsection{Lessons learned}
One of the missing features that should be implemented in such a system is a secure fast verification method. In early Bitcoin PoS designs, for each payment, the customer needs to wait 10 minutes in average for the transaction to be confirmed and included in the blockchain. We sidestep this issue by flagging the transactions as successful as soon as the transaction is broadcasted to the Bitcoin network, also known as a 0-confirmation transaction. This could work for a PoS in a caf\'{e} as the amount of each transaction is small and it is not significantly more risky to take 0-confirmation transactions than to risk a credit card chargeback or even a customer leaving the store without paying. Future work might consider PoS devices for the Bitcoin Lightening Network~\cite{poon2015bitcoin}. It remains an open problem to remedy the risk for higher value transactions and prevent double spend attacks\cite{karame2012two,bamert2013have}. 

Bitcoin and Bitcoin transactions are still new concepts for most people. We encountered a countless number of questions from customers to explain what Bitcoin is and how it works and mostly they became more interested to know more about Bitcoin when they observed a payment done with the Bitcoin PoS, mostly because they would not reveal any personal information with each payment.

Another interesting lesson is the concept of locked price that is the price of Bitcoin for each sale is locked to the exact exchange rate at the time of the transaction. This makes the acceptance of Bitcoin payments for the business risk-free, considering bitcoin to fiat conversion is done using the locked price in either monthly intervals or when a threshold is reached (\eg 100 dollars).

%% file: sections/newtable.tex

\begin{table*}[ht!]

\renewcommand{\arraystretch}{1.3}

\centering

\begin{tabular*}{0.9\textwidth}{@{\extracolsep{\fill}} llccccccccccccc}

\textit{Category} &
\headrow{User Friendly} & 
\headrow{Time-Efficient} &  
\headrow{Fair Exchange Rate} &
\headrow{Availability} &
\headrow{Low Cost to Run} &
\headrow{Enables Branching} & 
\headrow{Maintains Payee's Privacy} &
\headrow{Maintains Payer's Privacy} &
\headrow{Confidential Payments list} &
\headrow{No 3rd-Party Trust} & 
\headrow{Data Ecnryption} & 
\headrow{No Software Dependency} & 
\headrow{ } & 
\headrow{ } \\ \hline 

Displayed Address				&	&	&	&\full	&\full	&\prt	&	&\prt	&	&\full	&	&\full&&\\
Hardware Terminal 			&\prt	&\full	&\prt	&\full	&	&	&\full	&\prt	&\full	&	&\full	&	&&\\
Online Merchant Services	&\full	&\full&\prt	&\prt	&\prt	&\full	&\full	&\prt	&\prt	&	&\prt	&\full	&&\\ 
Mycelium Gear				&\prt	&\full	&\prt	&\prt	&\prt	&\full	&\full	&\prt	&\prt	&\prt	&\prt&\prt	&&\\ 
Aunja PoS				&\full	&\full	&\full	&\full	&\prt	&\prt	&\full	&\prt	&\full	&\full	&\full	&\prt	&&\\  \hline

\\
																					
\end{tabular*}

\caption{A comparison of Point of Sale gateways. \full~ indicates the category of client is awarded the benefit in the corresponding column. \prt~partially awards the benefit. Details provided inline.}
\label{tab:method-comp}

\end{table*}

%% file: main.bbl
\begin{thebibliography}{10}
\providecommand{\url}[1]{#1}
\csname url@samestyle\endcsname
\providecommand{\newblock}{\relax}
\providecommand{\bibinfo}[2]{#2}
\providecommand{\BIBentrySTDinterwordspacing}{\spaceskip=0pt\relax}
\providecommand{\BIBentryALTinterwordstretchfactor}{4}
\providecommand{\BIBentryALTinterwordspacing}{\spaceskip=\fontdimen2\font plus
\BIBentryALTinterwordstretchfactor\fontdimen3\font minus
  \fontdimen4\font\relax}
\providecommand{\BIBforeignlanguage}[2]{{%
\expandafter\ifx\csname l@#1\endcsname\relax
\typeout{** WARNING: IEEEtranS.bst: No hyphenation pattern has been}%
\typeout{** loaded for the language `#1'. Using the pattern for}%
\typeout{** the default language instead.}%
\else
\language=\csname l@#1\endcsname
\fi
#2}}
\providecommand{\BIBdecl}{\relax}
\BIBdecl

\bibitem{alford1979software}
M.~Alford, \emph{Software Requirements Engineering Methodology}.\hskip 1em plus
  0.5em minus 0.4em\relax Wiley Online Library, 1979.

\bibitem{androulaki2013evaluating}
E.~Androulaki, G.~O. Karame, M.~Roeschlin, T.~Scherer, and S.~Capkun,
  ``Evaluating user privacy in bitcoin,'' in \emph{Financial Cryptography and
  Data Security}.\hskip 1em plus 0.5em minus 0.4em\relax Springer, 2013, pp.
  34--51.

\bibitem{bamert2013have}
T.~Bamert, C.~Decker, L.~Elsen, R.~Wattenhofer, and S.~Welten, ``Have a snack,
  pay with bitcoins,'' in \emph{Peer-to-Peer Computing (P2P), 2013 IEEE
  Thirteenth International Conference on}.\hskip 1em plus 0.5em minus
  0.4em\relax IEEE, 2013, pp. 1--5.

\bibitem{BHOS12}
J.~Bonneau, C.~Herley, P.~C. van Oorschot, and F.~Stajano, ``The quest to
  replace passwords: a framework for comparative evaluation of web
  authentication schemes,'' in \emph{IEEE Symposium on Security and Privacy},
  2012.

\bibitem{BNMC+14}
J.~Bonneau, A.~Narayanan, A.~Miller, J.~Clark, J.~A. Kroll, and E.~W. Felten,
  ``Mixcoin: Anonymity for bitcoin with accountable mixes,'' in \emph{Financial
  Cryptography}, 2014.

\bibitem{Cha82}
D.~Chaum, ``{Blind signatures for untraceable payments},'' in \emph{{CRYPTO}},
  1982.

\bibitem{dorfman1990system}
M.~Dorfman, ``System and software requirements engineering,'' in \emph{IEEE
  Computer Society Press Tutorial}.\hskip 1em plus 0.5em minus 0.4em\relax
  Citeseer, 1990.

\bibitem{eskandari2015first}
S.~Eskandari, D.~Barrera, E.~Stobert, and J.~Clark, ``A first look at the
  usability of bitcoin key management,'' in \emph{Workshop on Usable Security
  (USEC)}, 2015.

\bibitem{karame2012two}
G.~Karame, E.~Androulaki, and S.~Capkun, ``Two bitcoins at the price of one?
  double-spending attacks on fast payments in bitcoin.'' \emph{IACR Cryptology
  ePrint Archive}, vol. 2012, p. 248, 2012.

\bibitem{MGGR13}
I.~Miers, C.~Garman, M.~Green, and A.~D. Rubin, ``{Zerocoin: Anonymous
  Distributed E-Cash from Bitcoin},'' in \emph{IEEE Symposium on Security and
  Privacy}, 2013.

\bibitem{Nak08}
S.~Nakamoto, ``Bitcoin: A peer-to-peer electionic cash system,'' Unpublished,
  2008.

\bibitem{bip32proposal}
{Pieter Wuille}, ``{BIP32 Hierarchical Deterministic Wallets},''
  \url{https://github.com/bitcoin/bips/blob/master/bip-0032.mediawiki}.

\bibitem{poon2015bitcoin}
J.~Poon and T.~Dryja, ``The bitcoin lightning network: Scalable off-chain
  instant payments,'' Technical Report (draft). https://lightning. network,
  Tech. Rep., 2015.

\bibitem{REScenario}
A.~Sutcliffe, ``Scenario-based requirements engineering,'' in
  \emph{Requirements Engineering Conference, 2003. Proceedings. 11th IEEE
  International}, Sept 2003, pp. 320--329.

\end{thebibliography}
